\documentclass{emulateapj}

\def\alwaysmath#1{\ifmmode{#1}\else{$#1$}\fi}

\def\deg{\circ}

\def\beq{\begin{equation}}
\def\eeq{\end{equation}}
\def\beqa{\begin{eqnarray}}
\def\eeqa{\end{eqnarray}}
\def\lsim{\lower0.6ex\vbox{\hbox{$ \buildrel{\textstyle <}\over{\sim}\ $}}}
\def\gsim{\lower0.6ex\vbox{\hbox{$ \buildrel{\textstyle >}\over{\sim}\ $}}}

\begin{document}

\title{The Effect of Substructure on Mass Estimates of Galaxies}
\author{Brian M. Yencho \altaffilmark{1}, Kathryn V. Johnston \altaffilmark{1}, James S. Bullock\altaffilmark{2} \& Katherine L. Rhode\altaffilmark{1,3}}
\altaffiltext{1}{Van Vleck Observatory, Wesleyan University, Middletown, CT 06459, 
USA; byencho@astro.wesleyan.edu, kvj@astro.wesleyan.edu, kathy@astro.wesleyan.edu}
\altaffiltext{2}{Center for Cosmology, Department of Physics \& Astronomy,
        University of California, Irvine, CA 92697, USA; bullock@uci.edu}
\altaffiltext{3}{NSF Astronomy \& Astrophysics Postdoctoral Fellow}

\begin{abstract} 
Large galaxies are thought to form hierarchically, from the accretion and disruption of many smaller galaxies.  Such a scenario should naturally lead to galactic  phase-space distributions containing some degree of substructure.  We examine the errors in mass estimates of galaxies and their dark halos made using the projected phase-space distribution of a tracer population (such as a globular cluster system or planetary nebulae) due to falsely assuming that the tracers are distributed randomly.  The level of this uncertainty is assessed by applying a standard mass estimator to samples drawn from 11 random realizations of galaxy halos containing levels of substructure consistent with current models of structure formation.  We find that substructure will distort our mass estimates by up to $\sim$20\% --- a  negligible error compared to statistical and measurement errors in current derivations of masses for our own and other galaxies.  However, this represents a fundamental limit to the accuracy of any future mass estimates  made under the assumption that the tracer population is distributed randomly, regardless of the size of the sample or the accuracy of the measurements.
\end{abstract}


\section{Introduction}
\label{section:introduction}

Measuring the masses of  galaxies is fundamental to understanding their nature. By the mid 1970's, observations \citep[e.g.][]{rubin80}  clearly indicated  that the luminous matter in galaxies that we can directly detect represents just a small fraction of the total mass present. This discovery was consistent with theoretical work which showed that the stability  of the Galactic disk necessitated the presence of a previously undetected ``massive halo''  \citep{ostriker73}, and before long, the now-standard picture of the luminous components of galaxies embedded deep within much more extended dark matter distributions was established.

Today, we have well-developed  ideas about the formation and evolution
of these extended dark  matter  halos --- the so-called  "hierarchical
structure formation" scenario  where small structures in  the Universe
form first   and   over   time   merge  to  form   larger   structures
\citep{peebles65,press74,blumenthal84}.   This   picture  successfully
explains    a    wide  range   of     observations  on  large   scales
\citep[e.g.][]{eisenstein05,maller05,tegmark04,spergel03,percival02},
but the evolution of    luminous  matter  in galaxies    within  these
structures is less well understood. For example,  in a recent analysis
of $10^5$  galaxies   taken from the  Sloan  Digital  Sky Survey,  the
mass-to-light ratios of galaxies was found to vary systematically with
their stellar mass  \citep[e.g.][]{kauffmann03}. Such trends represent
important constraints on the  general relationship between dark matter
halos and the luminous galaxies that live within them.

The trends in dark matter   content described above were derived  from
the central properties  of a large  sample of galaxies. An alternative
approach is  to use  tracer populations  (such as  globular  clusters,
planetary nebulae   or giant stars) to probe   large  distances from a
central galaxy,  allowing much more accurate  pictures of the mass and
extent of  the surrounding dark matter halo  to  emerge. This approach
derives mass estimates from the projected separation and line-of-sight
velocities of the tracer population.

A variety of such  mass estimators have been developed  but only a few
are commonly  applied today.  One  of   the earliest is  based on  the
virial theorem and  modern forms of it are  known  as the Virial  Mass
Estimator  \citep{zwicky33,limber60}.    Its formulation relies     on
independent averages of the   kinetic  and potential energies  of  the
system, resulting  in a  mass  estimate proportional  to ${ {  \langle
{v_{\rm los}}^2 \rangle } \slash{ \langle{ G  \over{R}} \rangle } } $,
where $v_{\rm los}$ refers to the line-of-sight velocity of an object,
$R$ is its  projected distance from the center  of the system,  $G$ is
the  gravitational constant and brackets  indicate an  average over an
entire  sample.   Other   estimators,  such  as  the    Projected Mass
Estimator, involve averages over  a quantity known as the  ``projected
mass'',  defined as   ${  (  v_{\rm{los}   }^{2}   R )  \slash{G}   }$
\citep{bahcall81,heisler85}.  Both these techniques give estimates for
self-gravitating systems and  are generally applied to galaxy clusters
and groups \citep{zwicky33,zwicky37,smith36,limber60}.

While these techniques have yielded estimates consistent with the existence of dark matter within galaxies, they can only be applied in cases where the number density of objects in the system is proportional to the underlying mass density.  This assumption is not likely to be true for tracer populations around galaxies such as globular clusters and planetary nebulae.  \citet{evans03} recently proposed a more general estimator that can be applied to these systems, where the densities of the light and mass follow different distributions: the Tracer Mass Estimator (hereafter TME).

Irrespective of the application, all of these estimators are derived assuming the samples that will be used are virialized and isotropic in spatial distribution and velocity space (or at least described by simple analytic functions).  At first it may seem that these assumptions should be valid for the case of spiral galaxies, since a basic model of a spiral galaxy halo is a spherically symmetric group of objects on random orbits.  Similarly, ellipticals may have some degree of flattening but these types of anisotropies could conceivably be corrected for when using the TME.  However, hierarchical structure formation suggests that these pictures of galactic structure may not be correct in detail. For example, observations of low velocity dispersions of stars in the around elliptical galaxies has recently been cited as evidence for lack of a dark matter halo \citep{romanowsky03}.  \citet{dekel05} show that such low dispersions arise naturally if ellipticals are formed predominantly from the merger of two spiral galaxies (consistent with the hierarchical scenario). They found that the orbits of stars at the end of simulations of the merger of two spiral galaxies tend to be  highly radial in the outskirts of the merger remnant. This radial anisotropy leads to  projected velocity dispersions much lower than expected at large radii, even in the presence of a substantial dark matter halo. 

The hierarchical scenario also naturally leads to structures in the Universe forming inside out \citep[i.e. with the inner parts forming first, see ---][]{helmi03,bullock05}, and in  sequence in scale, smallest to largest. It predicts that structures on the scale of galaxy clusters are still forming today, and hence that clusters are a poor place to assume virial equilibrium, as mass estimators do.  In contrast, the main epoch of galaxy formation has passed, and the bulk of galaxies should be virialized.  However, when estimating the mass of a galaxy, it is of most interest to find a tracer population that extends to large radii to probe far out into the dark matter halo.  It is precisely these regions on galactic scales that are still forming: even though the bulk of the galaxy may be virialized, incomplete mixing of debris from recent mergers (i.e. within last few Gyr) could lead to significant substructure in the phase-space distributions of the tracer population \citep[e.g.][]{johnston98,helmi99,bkw01}. Such substructure would violate the assumptions of spatial and velocity isotropy.

Spatial substructure has been observed in the outskirts of our own and other galaxies, lending strong support for the hierarchical structure formation model.  For example, when Ibata et al. (1994, 1995) discovered the Sagittarius dwarf galaxy it was clear that it is currently merging with the Milky Way.  While this work only hinted at ``tidal distortions'', later work by various groups found associated streams of debris extending all around the Galaxy \citep[see][for a summary]{majewski03}.  The tidal debris of other satellites of the Galaxy have also been mapped \citep[e.g. the Carina dwarf spheroidal][]{majewski00b} and similar features have been discovered in the halos of other galaxies \citep[e.g. M31][]{ibata01}.

As noted above, samples with spatial overdensities or correlated motions would break the assumption of a tracer population randomly distributed in phase-space.  For example, as globular clusters and giant stars can be used to discover tidal features \citep{lyndenbell95,palma02}, it is clear that they may not be distributed randomly, even in samples where spatial substructure is not immediately obvious.  \citet{wilkinson99} estimated quite how badly these broken assumptions might affect mass estimates  by comparing the result of applying a mass estimator to a truly random test data set with one in which all the data lies along  two streams. They concluded that there will be a 20-50\% error in the mass estimate in this particular case, considering it an upper limit since the substructure example they chose was so far from random. However, this work did not look in general at the expected level of substructure around galaxies today.

The goal of this paper is to assess the effect of substructure on mass estimates for galaxies formed hierarchically in a consistent cosmological context.
Rather than looking at specific, contrived  examples of substructure \citep[such as][]{wilkinson99}
we will draw samples from model stellar halos built entirely by accretion \citep{bullock05} and use the TME to provide mass estimates for these halos.  These samples were chosen to be similar to true observations of the different types of tracer populations, allowing us to assess the accuracy and precision of estimates that have been made and others that could be made in the future.

The TME and model datasets are discussed in greater detail in \S \ref{section:Methods}, while the samples taken from these datasets and the mass estimates provided from them are described in \S \ref{section:Results}. Our conclusions are summarized in \S 4.

\section{Methods}
\label{section:Methods}

\subsection{The Tracer Mass Estimator}
\label{section:TME}
To derive their estimator, \citet{evans03} assumed that the tracer populations and underlying mass distribution of a system can be approximated by certain standard forms.  The tracer population itself, gathered between some spherical radii $r_{\rm{in}}$ and $r_{\rm{out}}$, must have a number density that is spherically symmetric and described by a single power law in this region:
\begin{equation} 
\label{eq:density-TME}
\rho \left( r \right) = \rho_{\rm 0} \left( a \over r \right)^{\gamma} \; , \; r_{\rm in} \lesssim r \lesssim r_{\rm out}
\end{equation}
\noindent where $\rho_{\rm 0}$ is the density at $r = a$ or at all radii when $\gamma = 0$.  The underlying potential of the system is also characterized by a general form:
\begin{equation} 
\label{eq:potential-TME}
\Phi \left( r \right) = \left\{ \begin{array}{ll}
    {v_{\rm 0}^2 \over \alpha} {\left( b \over r \right)}^{\alpha} & \textrm{if $\alpha \ne 0$}\\
    \\
    {-v_{\rm 0}^2} \textrm{ log $r$} & \textrm{if $\alpha = 0$}\\
\end{array} \right.
\end{equation}
\noindent where $b$ is a constant with units of length and $v_{\rm 0}$ is the constant circular velocity of test particles when $\alpha = 0$.  The form of the potential for $\alpha = 0$ describes galaxies with flat rotation curves, for $\alpha = 1$ the potential is Keplerian, and $\alpha = \gamma - 2$ describes a self-gravitating, mass-follows-light system.  

Assuming a tracer population, with density distribution given by equation (1) and moving with an isotropic velocity distribution in a potential of the form given in equation (2), \citet{evans03} determined how the average value of the projected mass $\langle v_{\rm los}^2 R \rangle$ relates to the mass within the outermost radius
\begin{equation}
\label{eq:TME}
M( r_{\rm out} ) = {C \over G}{\langle v_{\rm los}^2 R \rangle} = {C \over G}{1 \over N}{\sum_{i=1}^{N} v_{{\rm los},i}^{2} R_{i}}
\end{equation}
\noindent where $N$ is the total number of objects in the sample and $C$ is defined as
\begin{equation}
\label{eq:C}
C =  \left\{ \begin{array}{ll}
{{4(\alpha + \gamma)} \over \pi} 
{{4 - \alpha - \gamma} \over {3 - \gamma}} 
{{1 - {(r_{\rm in}/r_{\rm out})}^{3 - \gamma}} \over {1 - {(r_{\rm in}/r_{\rm out})}^{4 - \alpha - \gamma}}} &
\textrm{if $\gamma \ne 3, \alpha+\gamma \ne 4$} \\  \\

{4 (\alpha+3)(1-\alpha) \over \pi}{\log(r_{\rm out}/r_{\rm in}) \over 1 - (r_{\rm in}/r_{\rm out})^{1-\alpha}} 
& \textrm{if $\gamma=3$}\\  \\

{16 \over \pi (3-\gamma)}{1 -(r_{\rm in}/r_{\rm out})^{3-\gamma} \over \log(r_{\rm out}/r_{\rm in})} 
& \textrm{if $\alpha+\gamma=4$}\\
\end{array} \right.
\end{equation}
\noindent These last two equations define the TME.  Note that while (\ref{eq:density-TME}) and (\ref{eq:potential-TME}) assumed the existence of the unknown constants $\rho_{\rm 0}$, $v_{\rm 0}$, $a$, and $b$ in order to derive the above equations, no knowledge of the true value of these constants is necessary in order to use this estimator.

\citet{evans03} also derived the TME for systems with anisotropic velocity distributions in which the ratio of the radial to the tangential velocity dispersion is allowed to vary.  We consider only the isotropic version since we expect the uncertainty due to substructure to be similar with either form of the estimator. Substructure represents a deviation from a random distribution that cannot be described by simply introducing anisotropy to the velocity distirbution.

\subsection{Discussion of Parameters}
\label{section:parameters}
The primary advantage of the TME is that it can be adjusted for different populations, giving it the widest possible application.  This is apparent in the four different parameters found in equation (\ref{eq:C}) for $C$: $\alpha$, $\gamma$, $r_{\rm in}$, and $r_{\rm out}$.  Since $C$ is the proportionality constant for the mass estimated from a given set of observational data, determining proper values for these parameters is vital to retrieving accurate estimates.

The parameter $\gamma$ is the power law fall-off of the density of the tracer population, and can be estimated directly from the tracer's surface density distribution in the region of interest (which should be chosen so that the tracer population is well-represented by a single value of $\gamma$).  Tests deriving $\gamma$ for different random samples drawn from the same population found variations in the estimates of order 12\% for a sample size of $N=50$, 3\% for $N=500$ and $<$1\% for $N=5000$.

The radii $r_{\rm in}$ and $r_{\rm out}$ represent the inner and outer spherical radii of the sample, and hence are unobservable.  However, since extended samples over a large range of radii are of most interest, and the density of tracer populations typically falls rapidly with radius (e.g. $\gamma\sim 3$ for the Milky Way globular cluster system), the spherical radii should be reasonably approximated by the projected inner $R_{\rm in}$ and outer $R_{\rm out}$ radii of the sample.

\begin{figure}[tb]
\plotone{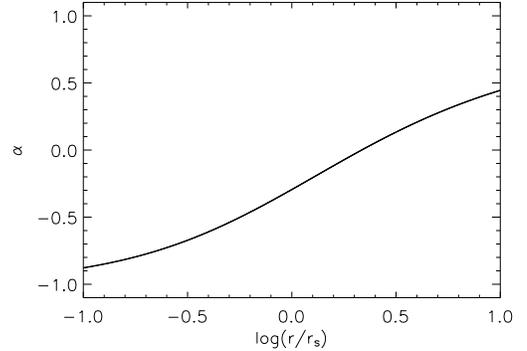}
\caption{\label{fig:alphaTest} The values of $\alpha$ corresponding to different values of $r \over r_{\rm s}$ for NFW halos, where $r_{\rm s}$ is the scale radius.}
\end{figure}

The most uncertain parameter is $\alpha$, which represents the unknown shape of the underlying potential (in fact, if the potential of the system were somehow known, then using a mass estimator would be irrelevant).  It is true that the behavior of the potential only needs to be known in the region probed by the tracer population, which \citet{evans03} always intended to be the outer regions of galaxies, where values of $\alpha \approx 0$ (corresponding to a flat rotation curve) should be appropriate.  \citep[Indeed,][ found that this approximation is valid out to nearly 170 kpc in the Milky Way.]{wilkinson99}  However, $N$-body simulations of structure formation suggest that dark matter halos in general should have potentials of the form
\begin{equation}
\label{eq:potential-NFW}
\Phi_{\rm halo}(r)  =  -{G M_{\rm halo}\over r_{\rm s}}{1 \over (r/r_{\rm s})} \ln\left({r \over r_{\rm s}}+1\right)
\end{equation}
\noindent \citep{navarro96,bullock01}, where $M_{\rm halo}$ and $r_{\rm s}$ are the mass and length scale of the halo respectively.  Figure \ref{fig:alphaTest} shows what values of $\alpha$ are expected for various regions of an NFW potential.  For  very large/small radii, the values of $\alpha$ tend toward $+1/-1$.  In practice, \citet{klypin02} estimate $r_{\rm s}$ in the range 15-30 kpc in order to fit observations of the Milky Way and Andromeda, so  our interest lies in the middle region, where $r$ is on the order of 0.1-10 $r_{\rm s}$ (i.e. few to hundreds of kpc).  No single value of $\alpha$ would seem appropriate here and, in fact, in the region around r $\sim r_{\rm s}$ $\alpha = -0.3$ rather than $\alpha = 0$.

\begin{figure}[tb]
\plotone{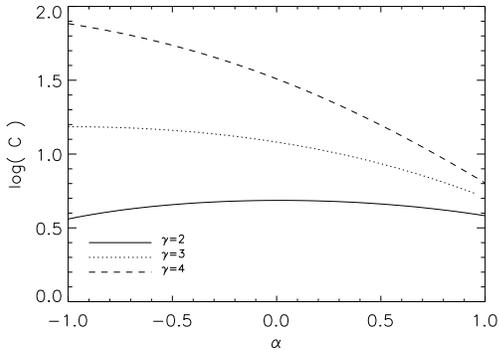}
\caption{\label{fig:C-Test1} $C$ as a function of $\alpha$ with $r_{\rm in} \slash r_{\rm out}$ = 20 for three different values of $\gamma$}
\end{figure}

In summary, the values of  $r_{\rm in}$, and $r_{\rm out}$ are known with some confidence because they can be estimated from the tracer sample itself.  The uncertainties in  $\gamma$ and $\alpha$ provide the main source of error in $C$ (and consequently the mass estimates).    Figure \ref{fig:C-Test1} shows $C$ as a function of $\alpha$ for three different values of $\gamma$.  The greatest range in $C$ 
occurs for the highest value of  $\gamma$ --- tracers whose density falls much more rapidly than the underlying mass distribution --- where large adjustments (i.e. large values of $C$) to the raw mass estimates are necessary.  
For an observed sample we might expect
$\gamma \sim 3$ (e.g., the globular cluster system of the Milky Way) and $\alpha$ to vary in the
range $-0.5$-$0.0$ in the region of interest (as discussed above and illustrated in Figure \ref{fig:alphaTest}). The range in $\alpha$ leads to an uncertainty in $C$ (and the mass estimate) of order 10-20\%. For small sample sizes ($N<50$), we showed that we will be unsure of our estimate of $\gamma$ at the 12\% level, which adds another $\sim 10$\% uncertainty to $C$. For larger samples, $\gamma$ will be better known and this contribution to the error budget will become negligible.

\subsection{Model data} 
\label{section:dataset}

\subsubsection{Description}
\label{section:description}

Our mock tracer population samples are drawn from the eleven stellar
halo models simulated by \citet{bullock05} (hereafter referred to as
Halos 1-11).  These models were created within the context of a
standard $\Lambda$ Cold Dark Matter universe, and hence are expected
to contain a degree of substructure representative of true galactic
halos.  \citet{bullock05} began with a Milky Way potential consisting
of three components: a halo, a disk, and a bulge.  Each of these
components grows over time according to analytic formulae.  More than
a hundred N-body simulations of dwarf galaxies disrupting within these
potentials are run for each halo, with the unbound particles
contributing to the growth of the halo.  The eleven models result from
eleven different randomly-realized accretion histories. It is these
histories that dictate the growth of the parent galaxy and the
accretion time and mass of each satellite.

Note that the stellar halo is represented in these models by assigning a variable mass-to-light ratio  to each dark matter particle in the simulations. These mass-to-light ratios are chosen so that the luminous matter in the infalling dwarfs initially follows a King model  embedded deep within a dark matter halo in the shape of an NFW profile. The samples described in \S \ref{section:samples}
are drawn at random from the full set of particles, weighted inversely by their mass-to-light ratios.

\citet{bullock05} confirmed the consistency of their models with the size and density profile of the Milky Way's stellar halo and the number and distribution in structural parameters of the Milky Way's satellite population. None of the resulting halos are perfectly smooth; they all contain some amount of substructure, which varies due to the different accretion histories.  In addition, each halo is somewhat flattened in response to the disk component of the parent galaxy potential.

\subsubsection{Isolating effects of substructure}
\label{section:isolatingSubstructure}
Although we are specifically interested in the effects of substructure, our models suffer from additional limitations that compromise the accuracy of the TME.  First, the flattening described in \S \ref{section:description} is a large scale deviation from isotropy that will clearly affect mass estimates: if the datasets are viewed ``face-on'' ($i = 0^{\deg}$) with respect to this flattening the velocity dispersion (and hence mass estimates) will be smaller than if they were viewed ``edge-on'' ($i = 90^{\deg}$).  Second, because the models are created by superposing individual simulations of satellite accretion rather than running a single, fully self-consistent simulation, we cannot expect them to be virialized. 

\begin{figure}
\plotone{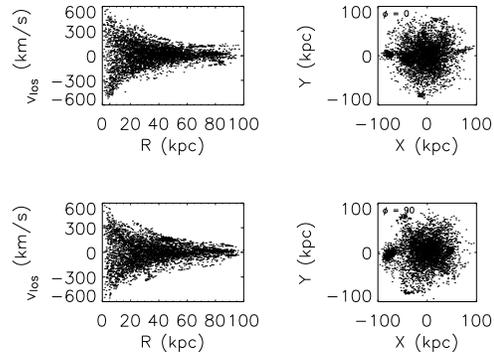}
\caption{\label{fig:sample-Substructure} The observable properties and projected positions of a 5000 particle sample taken from Halo 5 with an inclination of $90^{\deg}$.  The top row shows an azimuthal angle of $0^{\deg}$ while the bottom shows $90^{\deg}$. Differences in the location of substructure in the $v_{\rm los}$ {\it vs} $R$ panels suggest differences in mass estimates made using this data set.}
\end{figure}

The incorrect assumptions of spherical symmetry and virialization mean that we cannot expect the TME to recover the true mass of the underlying parent potential in our simulated data sets, even if substructure were not present.
Hence, rather than comparing to the true mass, we isolate the effects of substructure on our estimates for a given halo model by looking at how they vary along different lines of sight (i.e. azimuthal angle) all chosen at a fixed inclination to the plane of the parent galaxy disk.
 In a system with overdensities and streams, the correlated features would appear to move around the system (in $R$ and $v_{\rm los}$) when different angles are probed, as illustrated in Figure \ref{fig:sample-Substructure}, which will show up directly as a varying mass estimate.  

\section{Results}
\label{section:Results}
\subsection{Simulated samples}
\label{section:samples}
Samples are drawn from the test data set in two different ways to simulate: (i) an extra-galactic globular cluster survey \citep[i.e. that might be conducted today;][]{evans03}; and (ii) a large data set of giant stars (that might be collected in the future around Local Group galaxies).

\subsubsection{Simulating globular cluster surveys}

For this example, we assume that globular cluster systems are formed
primarily through the accretion of satellite galaxies and their
associated globulars.  This type of chaotic formation scenario has
been proposed by \citet{searle78}, who suggested that globular
clusters in the Galaxy's outer halo originated in infalling
protogalactic fragments.  Indeed, observations indicate that at least
some of the Milky Way globulars were likely to have been accreted from
smaller galaxies, since the Sagittarius dwarf is apparently
contributing four globulars to the overall Galactic system
\citep{ibata94}.  Furthermore, the properties of the globular clusters
in the Galaxy's outer halo --- such as their horizontal branch
morphologies, kinematics, and overall spatial distribution --- seem to
suggest that they were formed in satellite systems that were accreted
\citep{zinn93}.  On the other hand, a number of other scenarios have
been proposed for how globular cluster systems form (e.g., Forbes,
Brodie, \& Grillmair 1997, C\^ot\'e, Marzke, \& West 1998).  For
example, \citet{ashman92} theorized that a significant fraction of a
galaxy's globular clusters could be formed from dissipating gas during
galaxy-galaxy mergers, and observations of nearby merger events
indicate that massive star clusters do form in this way (Whitmore
2000).  Therefore the results in this section are likely to represent
an upper limit on the effect of substructure on mass estimates using
globular clusters as the tracer population.

\centerline{\it Selecting particles to represent globulars}

In order to select a realistic sample of $N$ globular clusters from our models an initial cutoff
is made such that all satellite galaxies with total luminosities less
than $10^{7} L_{\odot}$ are removed from consideration, as it would be
very unlikely that they would contain any globular clusters, which
themselves may have luminosities from $10^{4} L_{\odot}$ to $\sim
10^{6} L_{\odot}$. 
For the remaining satellites we have made the simplifying assumption
that the number of globular clusters is
proportional to the satellite's luminosity  --- i.e., the specific frequency
of globulars is constant.
In fact, the relationship between the properties of dwarf galaxies and the
number of globular clusters they host is not well-determined, and (for
example) the Local Group dwarf galaxies exhibit a wide range of
properties in terms of the relative number and metallicity range of
their globulars \citep{grebel02}.  

Next, given a line-of-sight to "view" the model
and the range of projected separations of the globulars from the parent (see below) 
all particles from satellites above our threshold luminosity value
within the specified range of separations are selected and saved.  Each
satellite is then assigned a sample size, which is a fraction of the
desired size $N$, in accordance with our specific frequency assumption. 
These samples are gathered from the individual satellites by designating
particles at random (weighted by the luminosity associated with each particle --- see \S 2.3.1) to be globular clusters.  Combining these gives
the final sample.

\begin{figure} [tb]
\plotone{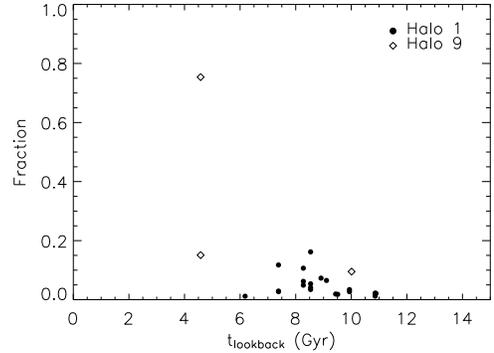}
\caption{\label{fig:globularFraction} The fraction of globulars assigned to the eligible satellites (i.e. satellites with luminosities greater than $10^7 L_\odot$) for two different halos, plotted as a function of the satellite's accretion time.}
\end{figure}

To illustrate the results of the selection process, 
Figure \ref{fig:globularFraction} shows how the luminosity, and
thus the globulars, will be distributed among all eligible satellites
in the $5 < R < 100$ kpc regions of two different halos chosen to be
at the extremes of the distribution of properties for all eleven
halos.  In these examples, Halo 1 has 16 satellites accreted $\sim$
8-10 Gyr ago contributing to this region with the most luminous
satellite containing about 16\% of the globulars.  In contrast, Halo 9
would have only 3 contributing satellites, 2 of which were accreted
4-5 Gyr ago, with the most luminous containing nearly 75\% of the
globulars.
  
\begin{figure} [tb]
\plotone{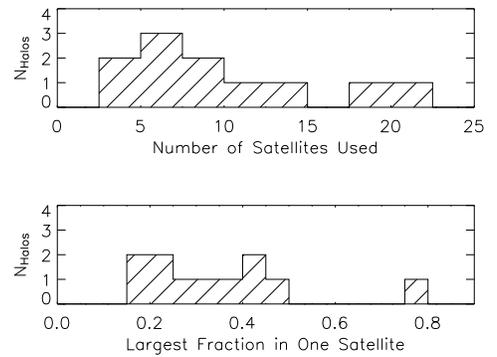}
\caption{\label{fig:numberAndFraction} 
The distribution of the number of satellites used (upper panel), and the largest fraction of globulars contributed by any one satellite (lower panel)  for all eleven halo models.}
\end{figure}

Given the extremes apparent in Figure \ref{fig:globularFraction}, Figure \ref{fig:numberAndFraction} summarizes the general properties
of infalling satellites for all halos by showing the distributions for
the number of eligible satellites and largest fractional luminosity
contributed to a halo by any single satellite.  The variety of
properties for the infalling satellite population evident in Figures
\ref{fig:globularFraction} and \ref{fig:numberAndFraction} is due to
the different accretion histories of each halo model.  Overall, we
expect the level of substructure in a halo (and hence a globular cluster system) 
to depend on how many
satellites contribute significantly to the halo's mass, how massive
the largest contributer is, and how far back the most recent merger
took place \citep[see][for a discussion]{bullock05}.  These figures
suggest that the TME may perform differently for the different models.

\begin{center}
{\it Number and extent of globular cluster samples in spiral and elliptical galaxies}
\end{center}

The range of radii around the
parent galaxy over which the sample will be taken and the number of
globulars that you would expect to be found in this range depends on the morphological
type of the parent.

Nearly half of the Milky Way's $\sim$ 180 globulars lie within 5 kpc,
while the most distant is located at around 100 kpc \citep{ashman98}.
In other spiral galaxies, most are located within around 20 kpc from
the centers of the host galaxies (Rhode, private communication).
Given these ranges, the systems in spirals are represented in our
datasets by samples of 50 globulars taken from 5 kpc to 50 kpc.

Elliptical galaxies typically have more populous globular cluster
systems than spiral galaxies.  Some giant ellipticals have globular
cluster systems with thousands of clusters, extending well beyond the
light of the galaxy (e.g., Ashman \& Zepf 1998, Rhode \& Zepf 2001,
2004).  For our simulations, we chose 500 particles over the radial
range of 5 to 100 kpc.  It should be noted here that the halo models
are specifically for spiral galaxies, not ellipticals; ellipticals are
likely to have undergone much more violent, recent mergers which could
effectively mix any phase-space substructure resulting from minor
mergers.  Thus, applying the results of this study to ellipticals will
give us an upper limit to the size of the error in the mass estimates
as it will overestimate the level of substructure existing in these
galaxies.  This example also serves to give us an indication of how
the uncertainty in the mass estimates due to substructure is related
to the sample size of a tracer population.

\subsubsection{Simulating giant star surveys}
Using systems of giant stars in extragalactic halos offers an alternative to using globular clusters. Such stars can be selected photometrically with some confidence \citep{majewski00a}.  Although the number of spectra are currently limited to a few hundred stars in fields probing a small fraction of the halos of Local Group galaxies, it is  feasible for such samples to be increased dramatically in both size and area coverage in the near future.
 
 In this case, there is no luminosity cutoff that needs to be made and no assumptions concerning the frequency with which these objects would be found in the satellite galaxies.  The particles are chosen from a range of 5 kpc to 100 kpc, which allows the resulting estimates to be compared to those from the samples described above.  A number of 5000 is chosen as the sample size.  While this is significantly larger than sample sizes of surveys conducted today, it is also much less than the total number of giant stars that exist in the halos of spiral galaxies.
 
\subsection{Mass estimates}
\label{section:results}
\subsubsection{Uncertainties due to substructure}

\begin{figure} [tb]
\centerline{\ }
\plotone{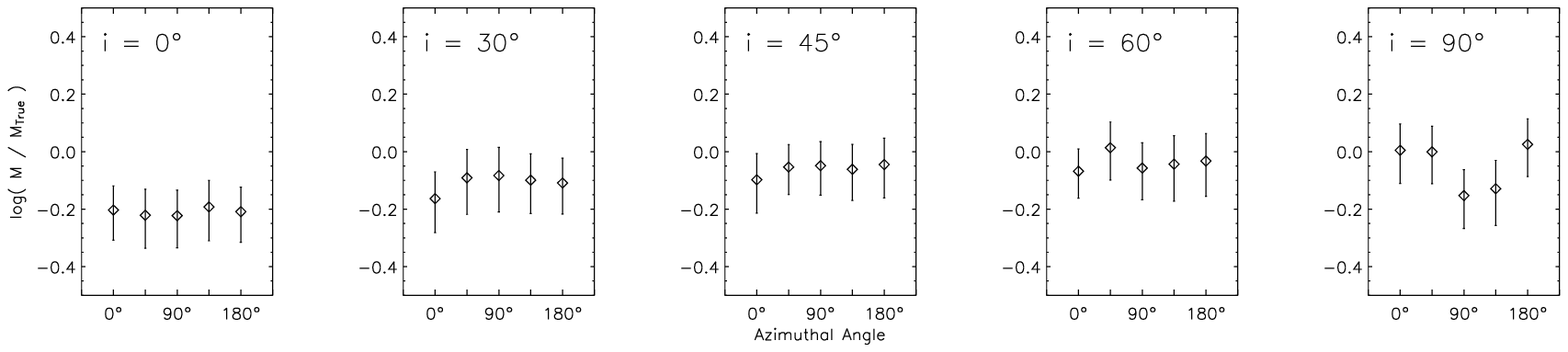}
\centerline{\ }
\centerline{\ }
\plotone{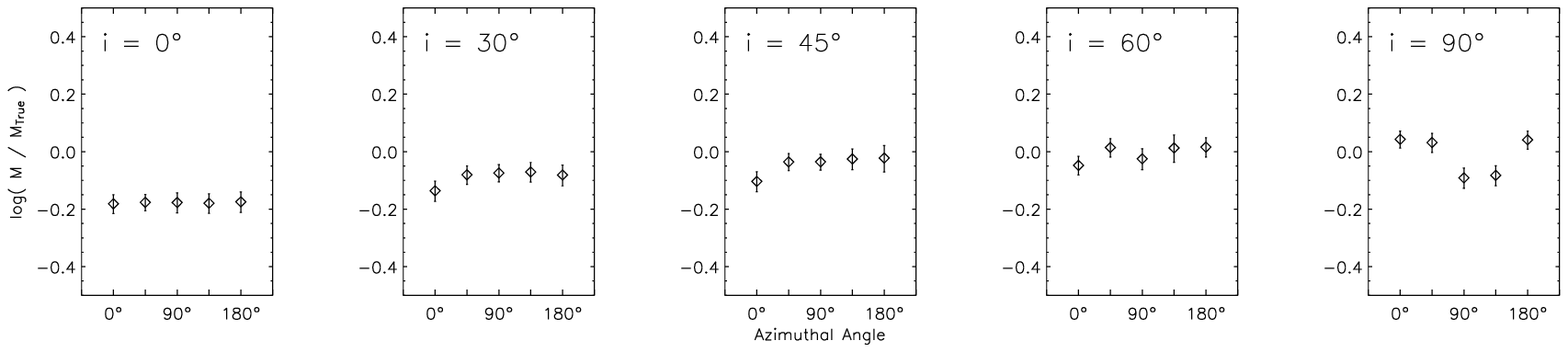}
\centerline{\ }
\centerline{\ }
\plotone{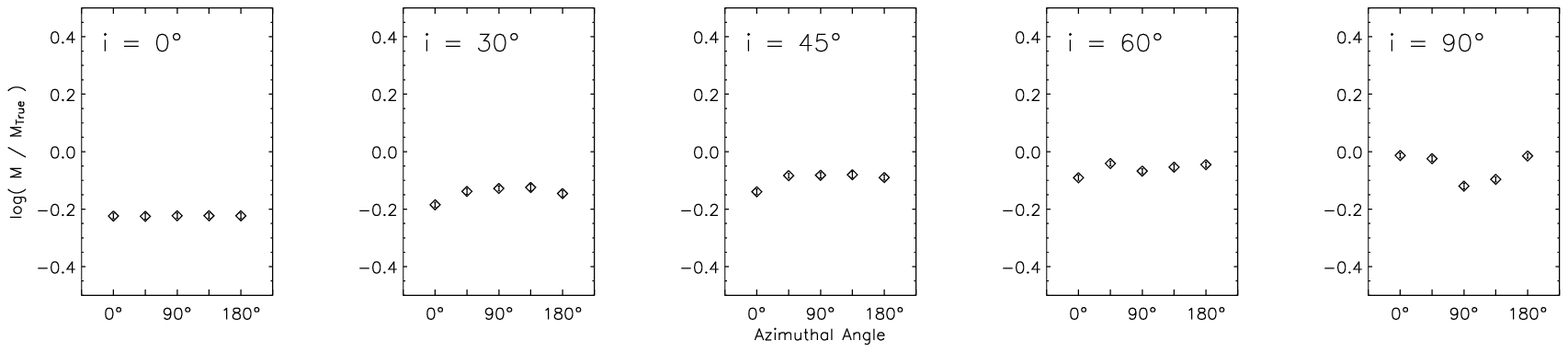}
\caption{ \label{fig:data-Halo1} Mass estimates for Halo 1 as a function of azimuthal angle and at the specified  inclination of the line of sight .  (Here, $i=0^o$ is face-on and $i=90^o$ is edge on.) The top row simulates a globular cluster survey of a spiral galaxy with $N=50$ and $5 < R < 50$ kpc.  The middle row simulates a globular cluster survey of an elliptical galaxy with $N=500$ and $5 < R < 100$ kpc.  The bottom row simulates a giant star survey with $N=5000$ and $5 < R < 100$ kpc.}
\end{figure}

Each panel of Figure \ref{fig:data-Halo1} shows the results of applying the TME to samples drawn from Halo 1 at fixed inclination (indicated in the panel) and as a function of azimuthal angle. (In this, and all subsequent figures, $M_{\rm True}$ is the known total mass in the simulation --- galaxy plus dark matter halo --- enclosed within the  outermost projected radius of the data sample.)
A value of $\alpha = 0$ was assumed for every sample, while $\gamma$ was calculated from the surface density of the data.  Each point represents the mean value of 100 different samples at the given inclination and azimuthal angle, while the error bars give one standard deviation.
As discussed in \S \ref{section:isolatingSubstructure}  we do not expect the average estimate to give a correct value even in the absence of substructure: the datasets are not virialized or spherically symmetric and $\alpha$ is unknown.
Rather, we interpret the scatter in the mean values in each panel as a measure of the effect of substructure.

The top row of panels in Figure \ref{fig:data-Halo1} shows the mass estimates for our
simulated globular cluster survey of a spiral (the most relevant sample to current observations).
The scatter in the mean lies well within the error bars for all points in this row (i.e. across both azimuthal and inclination angle). Hence the uncertainty in the estimate is clearly dominated by small number statistics. Moreover, our analysis excludes individual uncertainties in the observed positions and velocities, which have been shown to have a significant impact on the final uncertainty of the estimates \citep{wilkinson99}.    
These results generally hold for our full sample of 11 halos, although some halos have differences between values that are just beginning to reach a notable level of significance when comparing between inclinations of $0^{\deg}$ and $90^{\deg}$.  Overall, it is safe to say that for sample sizes associated with current observations, substructure does not contribute significantly to the error budget of mass estimates made with the TME.

The lower two panels of Figure \ref{fig:data-Halo1} demonstrate that variance in mass estimates due to substructure effects should become a dominant contributor to the error budget once sample sizes of order 500 or greater are obtainable.  It is interesting to note that the points (i.e. mean values of the mass estimates) in the middle and bottom panels fall at similar locations, which suggests similar results from our elliptical globular cluster  and giant star samples even though the former is restricted to being drawn only from debris from the more luminous satellites. 
In fact, $\sim$80\% of the mass in these stellar halos comes from the 15 most luminous satellites \citep{bullock05}, so the different sampling methods actually produce similar samples.

\begin{figure} [tb]
\centerline{\ }
\plotone{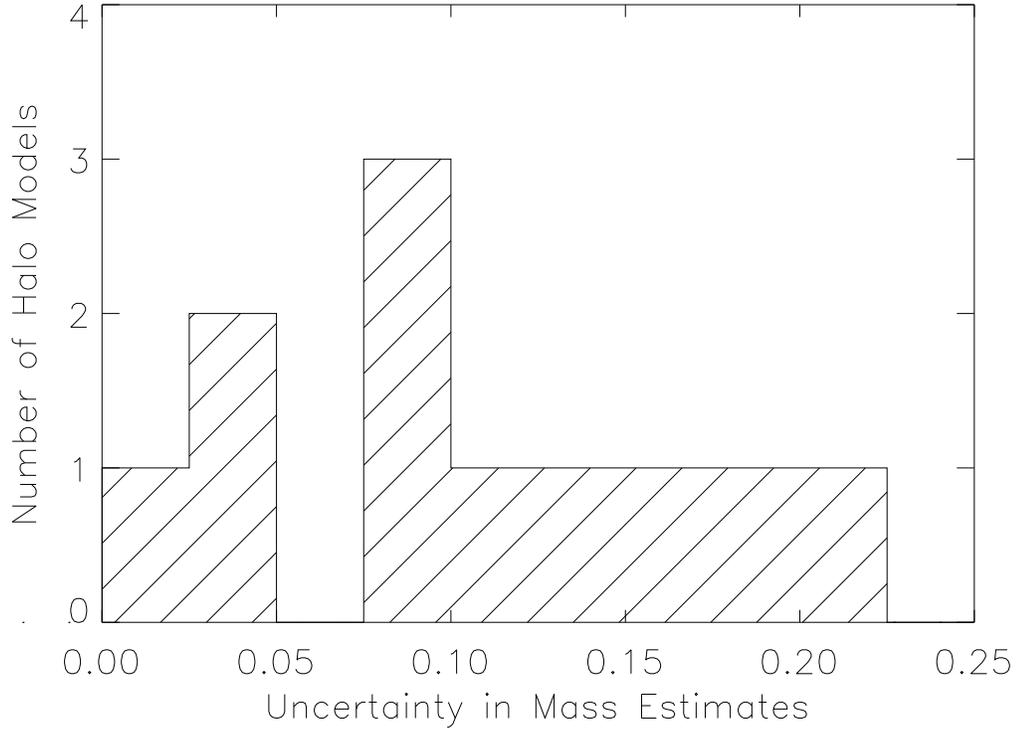}
\caption{ \label{fig:substructure} Histogram of the distribution of uncertainties due to substructure for all eleven halo models .  The mean of  the quantity $M_{\rm estimate} \slash M_{\rm true}$ was calculated from 10 $N = 5000$ samples at each of 51 azimuthal angles equally spaced in the range $0^{\deg}$ to $180^{\deg}$. The uncertainty was quantified by taking the standard deviation of the 51 means.}
\end{figure}

Figure \ref{fig:substructure} summarizes the results for all halos --- the histogram shows the distribution of the standard deviation of mass estimates made from samples of size $N=5000$ taken at 51 equally spaced azimuthal angles at an inclination of $90^{\deg}$ for each halo.    Based on this analysis, the effect of substructure can be quantified as introducing an uncertainty in mass estimates in the range 0-20\%, depending on the accretion history of the galaxy in question.  This represents a fundamental limit to the accuracy and precision of measuring the mass of a galaxy using the TME which can not be overcome by increasing the sample size.

\subsubsection{Uncertainties due to oblateness or triaxiality}

\begin{figure}
\plotone{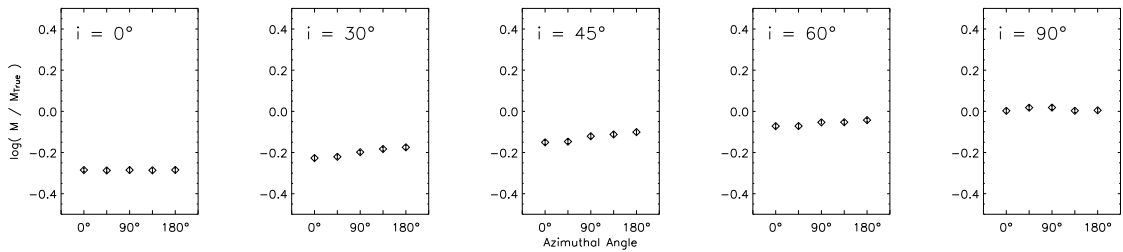}
\caption{ \label{fig:data-Halo3} Same as the bottom row of panels in Figure \ref{fig:data-Halo1} but for Halo 3.  The effect of inclination is seen in this case as a clearly increasing mass estimate. }
\end{figure}

Using the larger samples, it also becomes apparent that the intrinsic oblateness of the stellar halos causes the mass estimates to vary along different lines-of-sight. This is demonstrated in Figure \ref{fig:data-Halo3}, which shows the mass estimate as a function of inclination of the line of sight to the disk for our most flattened stellar halo, Halo 3 (minor-to-major axis ratio $c/a\sim 0.6$).  The points in each panel vary little in azimuthal angle (i.e. due to substructure), but the average mass estimated at an inclination of $90^{\deg}$ is nearly twice that of the average mass given for an inclination of $0^{\deg}$.  

If we can assume that the stellar halo of a spiral galaxy is purely oblate and that its symmetry plane is aligned with the disk, the systematic uncertainty in the mass estimate due to inclination to the line-of-sight could be accounted for by measuring the apparent flattening of the tracer population and correcting by the known inclination of the disk.   These assumptions are not too unreasonable. Although dark matter halos that form in cosmological N-body simulations tend to be triaxial, with minor-to-major axis ratios of order $c/a\sim0.6$   \citep{jing02,bullock02,allgood05}, simulations with a baryonic component, including the effects of  gas cooling and star formation lead to only mildly triaxial halos \citep[$c/a\sim 0.8$ and $b/a \sim 0.9$, see][]{kazantzidis04,kazantzidis05,bailin05} containing disks whose rotation axes align with the halo's minor axis \citep{bailin05}. Hence, after correcting by the disk inclination, the remaining asymmetry should be very mild ($b/a\sim 0.9$). Our own results (i.e. factor two variance in the mass estimate for $c/a\sim 0.6$) suggest that this would lead to an uncertainty in mass estimates of order 10\%.


\section{Conclusions}
\label{section:discussionAndConclusions}

This paper presented an analysis of errors inherent to the Tracer Mass Estimator, the most appropriate estimator to use when calculating the mass of a single galaxy using tracer population kinematics.  It was applied to samples drawn from 11 halo models which contained levels of substructure consistent with the current $\Lambda$-CDM paradigm of galaxy formation.  
Each of the following were found to contribute to the error budget for the 
mass estimates at the 10\% level: (i) incorrectly assuming
a scale-free power law potential throughout the region sampled; 
(ii) uncertainty in the measurement of the surface density distribution of the tracer population;
(iii) oblateness or triaxiality of the tracer population; and (iv) substructure in the tracer population.
(In our eleven examples, substructure caused misestimates in masses of up to $\sim$20\%.)

This level of systematic uncertainties is similar to that expected for mass estimates of galaxy clusters using intergalactic planetary nebulae. In her analysis of a numerical model of a galaxy cluster, \citet{willman04} found that applying the Projected Mass Estimator (PME) to unbound particles at the end of the simulation led to an overestimate of the cluster mass by of order 10\%. Her study suggests that this uncertainty was due to a combination of anisotropies in the velocity distribution of the tracer population with known limitations of the PME, rather than substructure. However, since the study was limited to one galaxy cluster,  it could not address the effect of substructure more generally.

Systematic errors of up to $\sim$20\% are not the main contributors to the error budget when using smaller sample sizes consistent with current surveys of spiral galaxy globular cluster systems ($N < 100$). In these cases the statistical and measurement errors dominate. However, once sample sizes of order $N=500$ are achieved (either galaxies with larger globular cluster systems or future samples of giant stars in nearby galactic halos), these systematic errors represent a fundamental limit of the TME.

Of course, a larger sample size can be utilized to refine the TME to solve for additional global properties, such as the full density profile of the underlying mass distribution (rather than assuming a single power law) and simple asymmetries in the tracer population distribution (i.e. to correct systematic errors (i), (ii) and (iii) above). Moreover, in the case of the Milky Way, future astrometric missions such as NASA's Space Interferometry Mission and ESA's Global Astrometric Interferometer for Astrophysics promise additional dimensions of phase-space information at very large distance from the Galactic center so that phase-space anisotropies could be directly assessed.
However, substructure in the tracer population is not something that can be simply parameterized and hence our study suggests a fundamental limit of 20\% to the accuracy of any mass estimator that is developed under the assumption of randomness, however large the sample or  accurate and complete the measurements.

\acknowledgments 
KVJ's contribution was supported through NASA grant
NAG5-9064 and NSF CAREER award AST-0133617.  KLR is supported by an
NSF Astronomy \& Astrophysics Postdoctoral Fellowship under award
AST-0302095.  JSB is supported by the Center for Cosmology at UC Irvine.

\normalsize


\end{document}